Response to "DISCOVERY OF THE ELEMENT WITH ATOMIC NUMBER 112" by ROBERT C. BARBER, HEINZ W. GÄGGELER, PAUL J. KAROL, HIROMICHI NAKAHARA, EMANUELE VARDACI and ERICH VOGT (PAC-REP-08-03-05)[1]

AMNON MARINOV[1], DIETMAR KOLB[2] and JESSE L. WEIL[3]

[1] *Racah Institute of Physics, The Hebrew University of Jerusalem, Jerusalem 91904, Israel*
[2] *Department of Physics, University GH Kassel, 34109 Kassel, Germany*
[3] *Department of Physics, University of Kentucky, Lexington, 40506, KY, USA, and Institute of Isotopes, Hungarian Academy of Sciences, POB 77, H-1525 Budapest, Hungary*

*Abstract:* Based on two α-particle chains the Joint Working Party (JWP) Report assigns the priority for discovering element 112 to work done in 1996 and 2002 at GSI, Darmstadt. By doing this it ignores the data presented to the JWP about the discovery of this element already back in 1971 by Marinov *et al.* In those measurements about one hundred fission fragments were observed from two Hg sources separated from two W targets irradiated with 24 GeV protons. Furthermore, the masses of the fissioning species were measured and interpreted as being due to the atom and four different molecules of element 112 with atomic mass numbers 272-273. By ignoring these data the JWP disregards the facts that mass measurement is considered by the Transfermium Working Group (TWG) report as an "excellent criterion" and that "chemical methods can yield excellent assignment criteria". It is shown below that all the arguments given by the JWP attempting to justify their decision are refutable. In particular, their refusal to accept the possibility of the existence of long-lived isomeric states as a consistent interpretation of the data is unjustified, since, among other observations, evidence has been found recently for the existence of such states in naturally-occurring neutron deficient $^{211,213,217,218}$Th isotopes, and in the superheavy element region, at atomic mass numbers A = 261 and A = 265 (most probably $^{261}$Rg and $^{265}$Rg), and at A = 292, Z ≅ 122 (eka-Th), with half-lives $t_{1/2} \geq 10^8$ y.

**a) INTRODUCTION**

Back in 1971 evidence was obtained for the production of element 112 via secondary reactions in W targets irradiated with 24 GeV protons [1,2]. About a hundred fission fragments were observed in two Hg sources separated from two W targets. In addition, the masses of the fissioning species were measured and high masses like 272, 288, 308, 315 and 317-318 were observed and consistently interpreted as the atom and four different molecules of element 112 with A = 272-273 [3]. In a parallel study of actinide nuclei produced in the same W target, evidence for the existence of long-lived isomeric states in the neutron deficient $^{236}$Am and $^{236}$Bk nuclei was observed [4]. At the time, the data were not understood. There was the long lifetime of several weeks for the fission activity and the large deduced fusion cross section of several mb that seemed to contradict established knowledge. Based on the observed isomeric states in $^{236}$Am and $^{236}$Bk, it was suggested [3] that, like in the actinides, a long-lived isomeric state has been produced in element 112. Regarding the large fusion cross section it was first argued that, from the experimental point of view if $10^4$ - $10^5$ atoms of $^{236}$Bk and $^{236}$Am were produced in the reaction, then the production of about 500 atoms of element 112 would not be impossible. Then it was pointed out [3] that the projectiles in secondary reaction experiments are not nuclei in their g.s., but rather highly excited deformed fragments produced within $5 \times 10^{-14}$ s before interacting with another W nucleus in the target. Deformations affect greatly the fusion cross sections as was demonstrated by the well known sub-barrier fusion effect [5,6], where an increase of four to five orders of magnitude is expected. At that time, about four to five orders of magnitude were still missing to fully understand the large fusion cross section.

---

[1] Preprint: submitted for publication to Pure and Applied Chemistry/

These data were evaluated by the TWG of IUPAC and IUPAP who issued their reports in 1991 and 1992 [7,8]. Generally speaking, the attitude of the TWG towards our work was favorable. A member of this group M. Lefort, wrote in a letter of March 20, 1991 [9] to A. Marinov that "The possibility of producing long lived isomeric states in neutron deficient very heavy nuclei is indeed reasonable, as well as the hypothesis of production of highly excited deformed fragments as possible projectiles." In addition, Professor Lefort pointed out that we had found unexpected results, "In your beautiful work which has been carried on during so many years, you have found, I think, many unexpected results…" [9]. In the minutes of the TWG meeting in Mogilany-Krakow on 1-5 July, 1991 they wrote about element 112: "The TWG is not yet convinced that this element was indeed seen, but it judges the experiments so interesting that it will express hope that they will get a follow-up." In its assessment it is written: "Little is known about the production of heavy ions such as Sr in such bombardment and even less about their energy distribution. Although it would be surprising if their yield were sufficient in the rather narrow energy band that would be operative for reactions such as that required for the production of element 112, this possibility cannot be definitively dismissed and further work is to be hoped for." [8].

The work that was hoped for by the TWG has been carried out by us and has yielded the discovery of new kinds of isomeric states, namely high spin isomeric states in the superdeformed (SD) and the hyperdeformed (HD) minima of the potential energy surfaces of the nuclei [10-13]. It has been shown [13] that the fusion cross sections for production of such isomeric states, where the shape of the compound nucleus is similar to the shape of the projectile-target combination at the touching point, are much larger than for producing the normal g.s. Thus, the combined effect of highly excited deformed projectiles (fragments) and the possible production of the compound nuclei in SD or HD isomeric states, resolves the fusion cross section problem in the secondary reaction experiment. It was also shown that the possible existence of such isomeric states provides a consistent interpretation for previously unexplained radioactive phenomena seen in natural materials, like the Po halos and the low energy 4.5 MeV α-particle group attributed to Hs [14,15]. Furthermore, accurate mass measurements were performed on natural Th and Au materials. Long-lived isomeric states with half-lives $t_{1/2} \geq 10^8$ y have been found in the neutron-deficient $^{211,213,217,218}$Th nuclei [16] and in the superheavy element region at masses 261 and 265, most probably $^{261}$Rg and $^{265}$Rg (eka-Au) [17], and at A = 292, Z $\cong$ 122 (eka-Th) [18].

All these data were submitted to the JWP. It was argued that since there is an obvious explanation for the fusion cross section problem, and furthermore that long-lived isomeric states with half-lives of $t_{1/2} \geq 10^8$ y exist in natural materials, there is no reason to ignore the original observation back in 1971 of element 112 with a half-life of several weeks. It seems that it was the first evidence for a more general phenomenon of long-lived high spin isomeric states in the SD and HD minima of the potential energy surfaces of the nuclei with unusual radioactive decay properties. Contrary to the TWG, the attitude of the JWP towards our works was completely negative. They basically rejected all our data. If they had accepted just one piece of evidence, for instance the observation of the very high energy and long-lived 8.6 MeV α-particle group in coincidence with SD band γ-rays (see Figs. 1 and 2), they would have had to recognize that long-lived SD and HD isomeric states do exist. This logically should have led them to a different conclusion about the discovery of element 112. Below we refute the JWP arguments in some detail.

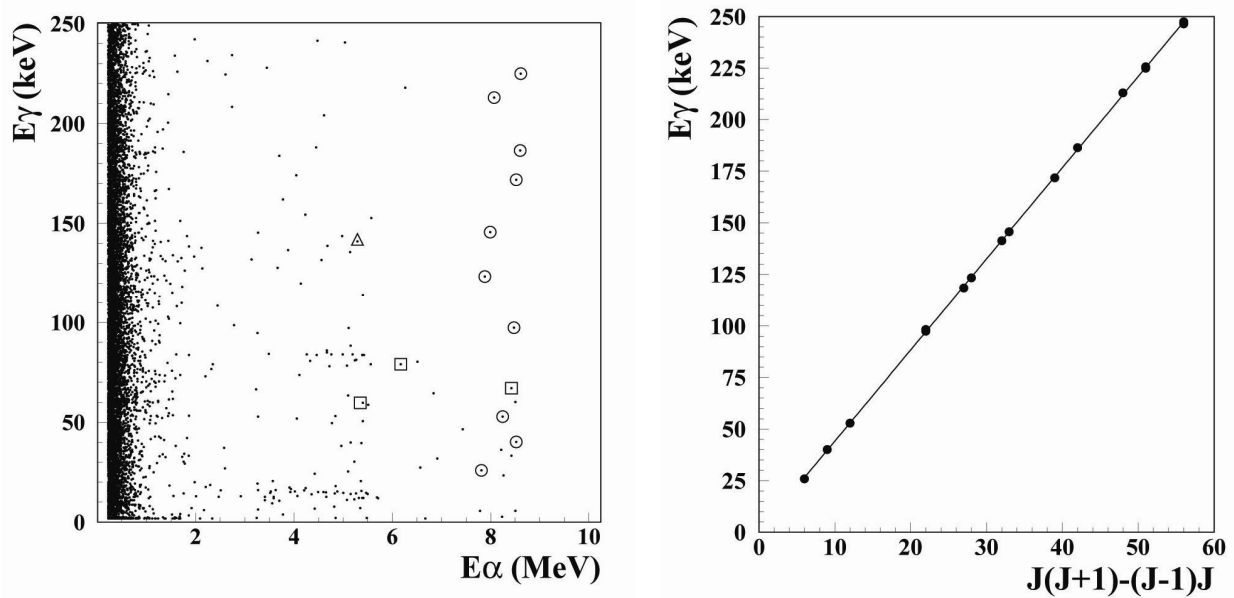

Fig. 1 (Left) (Fig. 5 of Ref. [12]) A two-dimensional plot of particle-γ-ray coincidences observed in the $^{28}$Si + $^{181}$Ta reaction. The γ-ray energies of the encircled events fit with SD band transitions.

Fig. 2 (Right) (Fig. 11 of Ref. [12]) $E_\gamma$ versus $J(J+1) - (J-1)J$ for the γ-rays seen in coincidence with 7.8-8.6 MeV α-particles (Figs. 5 and 6 of Ref. [12]).

**b) COMPARISON OF HOFMANN *et al.* and MARINOV *et al.* DATA**

On the basis of two time-correlated α-particle chains with time differences from about 100 μs to about 20 s, one of which was found in 1996 and the second one in 2002, the JWP awarded priority to the discovery of element 112 to the Hofmann *et al.* collaboration. One of these chains ended at a previously identified isotope of No, providing a sound basis for inferring the Z-value of the first element of the chain. But in awarding priority the JWP did not give proper consideration to all the evidence submitted to them on the discovery of this element back in 1971. This evidence is:

b1) All measurements were made on material following the chemistry of Hg [1].

PDF created with pdfFactory trial version www.pdffactory.com

b2) Observation of about a hundred fission events in Hg sources separated from two W targets irradiated with 24 GeV protons [1,2].

b3) A mass measurement of the fissioning material was performed and eleven fission events were subsequently observed at the positions corresponding to masses of 272, 288, 308, 315, and 317-318 (Fig. 3 (Fig. 2 of Ref. [3])). These masses were interpreted as the atom and four different molecules of an isotope of element 112 with mass number A = 272 or 273 (Table 1; Table 1 of Ref. [3]). The sample placed in the mass separator was a thin Cu foil that had been placed in an HCl solution of the Hg source obtained from the W2 target, allowing for electroplating of Hg and eka-Hg without applying voltage. This thin Cu foil was heated in the ion source of the mass separator to about 300° C, a temperature that excluded detecting any superheavy element below eka-Hg.

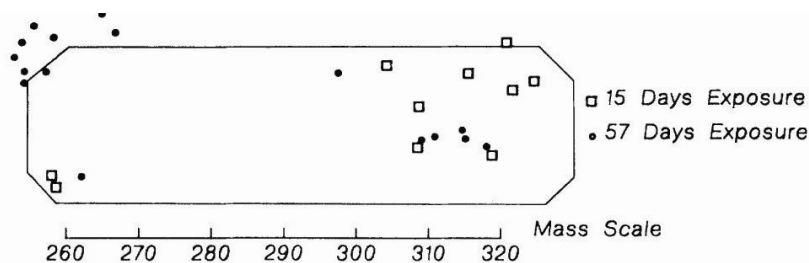

Fig. 3 (Fig. 2 of Ref. [3]. Copyright (1984) by the American Physical Society.) Position of fission fragments observed in the Ni foil by use of polycarbonate detectors. The octagonal form outlines the position of the Ni foil. The distance between mass units was 2 mm. (The points on the left-hand side are due to edge effects of the polycarbonate foils.)

TABLE 1 (Table 1 of Ref. [3]. Copyright (1984) by the American Physical Society.) Results of mass separator measurements on the Hg(W2) source. The number of fission tracks are given in parentheses for each mass. The masses are arranged according to various possible molecules of element 112 (see text).

| $A^+$ | $A^{16}O^+$ | $A^{35}Cl^+$ | $A^{12}C^{14}N^{16}O^+$, $A^{14}N_3^+$ | $A^{14}N^{16}O_2^+$ | $N$ |
|---|---|---|---|---|---|
| 269(1) | | | | | 157 |
| 272(1) | 288(1) | 308(3)[a] | 315(2) | 317–318(4) | 160–161 |
| 276(1) | 292(1) | 311(1)[b] | | | 164 |

[a] Mass 308 may also be interpreted as $^{276}AO_2^+$.
[b] Mass 311 may also be interpreted as $^{269}AN_3^+$ or $^{269}A^{12}C^{14}N^{16}O^+$.

One needs good arguments to justify ignoring these data. Such arguments are not presented in the JWP Report.

The JWP claims that observation of fission fragments is "a very nonspecific indicant" and uses this argument to dismiss our observation of fission activity as evidence of element 112 production.

There is a big difference between the observation of fission fragments on-line in ordinary heavy-ion reactions, where no other property like mass or chemical behavior is measured, and the measurements we performed in the secondary reaction experiments. In these experiments all the fission events were seen in species that followed the specific chemistry of Hg. For some of the events, this included the fact that the source material electroplated on Cu without applying voltage (a property of Hg) and evaporated at the low temperature of 300º C, and finally the fission events were found at definite measured high masses which were consistently interpreted as the atom and four different molecules of element 112 with A = 272-273. Identification of an element and a particular isotope of it by its chemical properties and its measured atomic mass number is a specific identification. The JWP should have said why they ignore these results when mass measurement is considered by *Criteria…*[7] as an "excellent criterion" [7] and "chemical methods can yield excellent assignment criteria" [7].

The JWP arguments against the results of the secondary reaction experiments are mainly concerned with the theoretical interpretation of the data and not with the data itself. Most of the arguments were already given in their previous reports [19,20] and were comprehensively answered in [21]. Let us mention briefly the JWP arguments and provide answers to them. As seen below all their arguments are refutable.

**c) CROSS SECTIONS IN SECONDARY REACTIONS**

The JWP repeated its argument about the cross section. "Unusually high fusion cross sections are required for nuclide formation, each several orders of magnitude beyond known behavior."

The answer to this question was given in Ref. [21] and detailed estimates of the cross sections involved in the secondary reaction production of both the various actinide nuclei and element 112 are given in Table 4 of Ref. [13]. As was mentioned in Ref. [13] it was estimated that only about $5 \times 10^{-5}$ of the produced fragments like $^{86,88}$Sr ions can have enough kinetic energy to be effective in the second step of the reaction. Hence, the deduced fusion cross sections are in the region of a few mb. As mentioned in the Introduction and in detail in Ref. [13], these large fusion cross sections in the actinides and element 112 are explained as being due to two effects (we briefly repeat the arguments):

c1) The projectiles in the secondary reaction experiments are not nuclei in their g.s. They are highly excited deformed fragments produced by a proton interaction with a W nucleus about $5 \times 10^{-14}$ s before interacting with another W nucleus in the target. Deformations have a strong effect on the fusion cross section between heavy nuclei, as demonstrated by the sub-barrier fusion effect [5,6] and seen in Fig. 4 (Fig. 5 of Ref. [6]. See also Fig. 7 of Ref. [13]). This argument was published in 1984 [3].

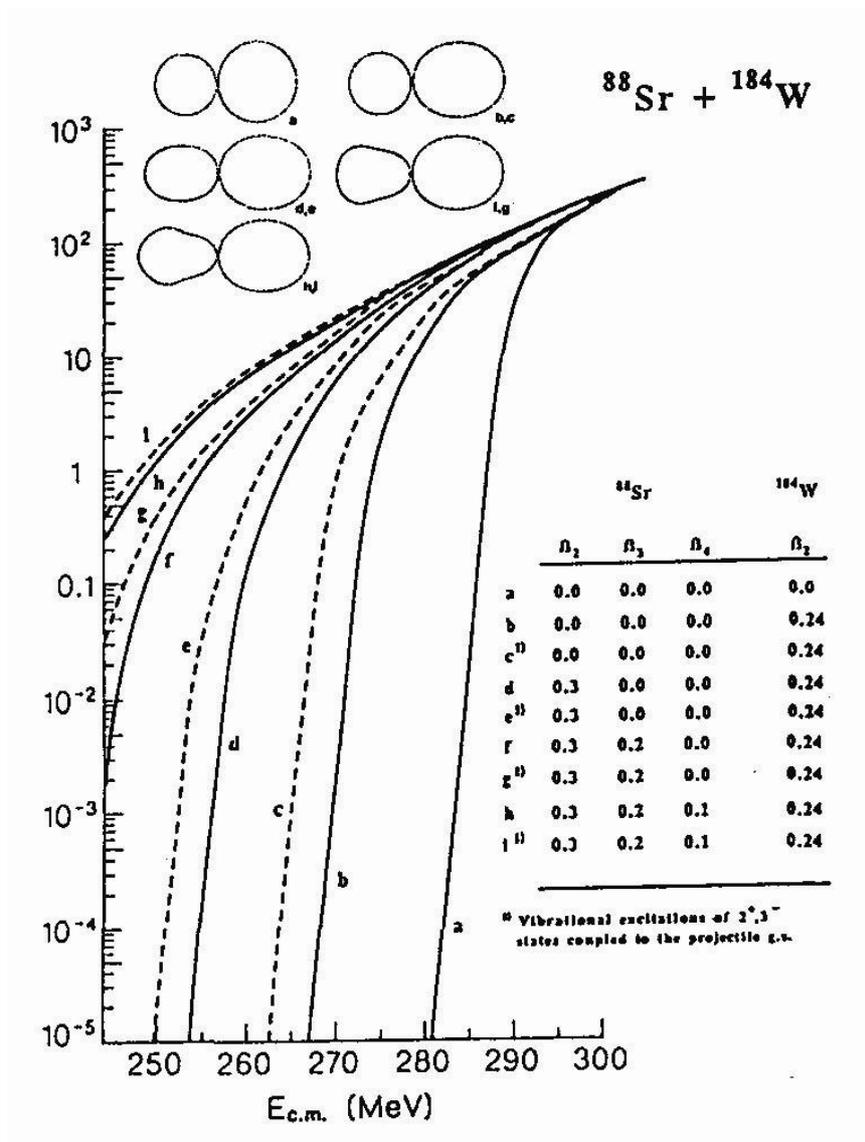

Fig. 4. (Fig. 5 of Ref. [6]). Calculated fusion cross sections using the code CCDEF for the $^{88}$Sr + $^{184}$W system assuming different deformations for the $^{88}$Sr projectiles (fragments).

c2) A second effect that can increase drastically the fusion cross section is the possible production of the compound nucleus in a SD or HD isomeric state, as seen clearly in Fig. 5 (Fig. 8 of Ref. [13]). Much less penetration and dissipation are needed when the shape of the compound nucleus is similar to the shape of the projectile-target combination at the touching point.

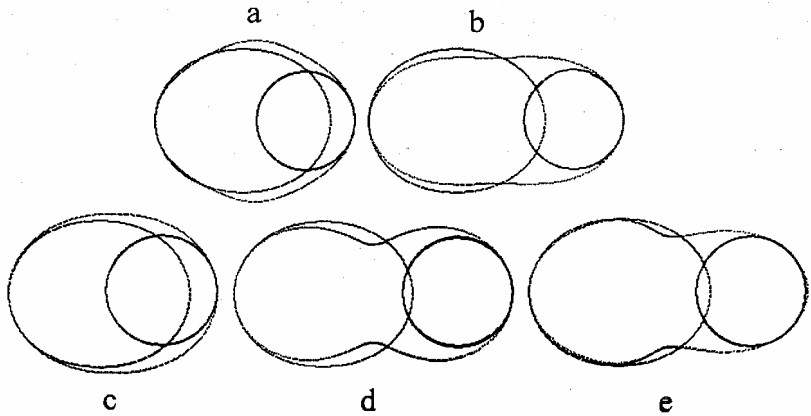

a) $A_{C.N.} = 239$; $\beta_2 = 0.2$; $\beta_4 = 0.08$.    b) $A_{C.N.} = 239$; $\beta_2 = 0.77$; $\beta_4 = 0.1$.
   $A_{heavy} = 186$; $\beta_2 = 0.22$. $A_{light.} = 53$; $\beta_2, \beta_3, \beta_4 = 0.0$.

c) $A_{C.N.} = 253$; $\beta_2 = 0.28$; $\beta_4 = 0.01$.    d) $A_{C.N.} = 253$; $\beta_2 = 1.2$; $\beta_4 = 0.0$.
e) $A_{C.N.} = 253$; $\beta_2 = 0.85$; $\beta_3 = 0.35$; $\beta_4 = 0.18$.
   $A_{heavy} = 186$; $\beta_2 = 0.22$. $A_{light.} = 67$; $\beta_2, \beta_3, \beta_4 = 0.0$.

Fig. 5 (Fig. 8 of Ref. [13]). Calculated shapes of two compound nuclei at various configurations together with the shapes of the corresponding projectile and target nuclei:

We list below the experimental evidence for the existence of long-lived isomeric states in neutron deficient heavy nuclei, and we specify those cases where their character as high spin SD and HD isomers was measured.

It is not understandable why the JWP insists on ignoring the chemical, mass and fission results of Sect. (b) above, when there is such an obvious explanation for the fusion cross section problem.[2]

---

[2] We would like to comment that the works of Brandt et al. [22] about unusually high cross sections for secondary particles in thick targets are irrelevant to our measurements. In their cases the secondary particles are projectile-like fragments and in our case fission-like nuclei as Sr are considered. All our estimates are based on experimental data [13].

**d) ISOMERIC STATES**

As mentioned in the Introduction, when Professor Lefort of the TWG wrote in 1991 [9] that: "The possibility of producing long-lived isomeric states in neutron deficient very heavy nuclei is indeed reasonable…", the only long-lived isomeric states found by us were in $^{236}$Am and $^{236}$Bk [4]. Since then many more isomeric states have been found. Let us summarize in more detail the experiments where long-lived isomeric states were seen and where their structure as high spin SD and/or HD states were measured.

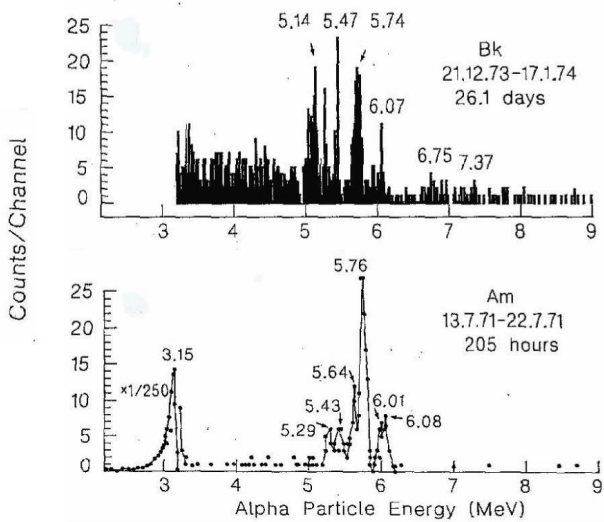

Fig. 6 (Part of Fig. 1 of Ref. [4]. Copyright (1987), with permission from Elsevier.) α-particle spectra from Am and Bk sources separated from the W(3) target irradiated with 24 GeV protons.

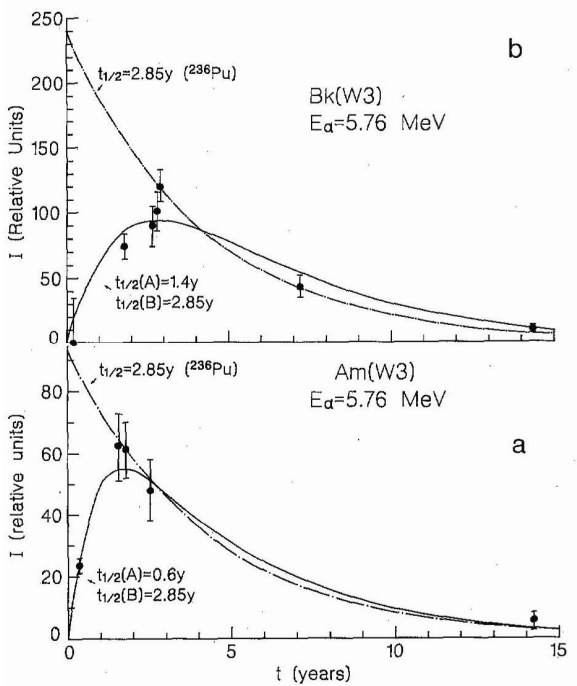

Fig. 7 (Fig. 2 of Ref. [4]. Copyright (1987), with permission from Elsevier.) Decay curves for the 5.76 MeV α-particle group seen in the Am (a) and Bk (b) sources as demonstrated in Fig. 6.

d1) In the secondary reaction experiments a 5.76 MeV α-particle group has been seen in chemically separated Am and Bk sources (See Fig. 6 (Part of Fig. 1 of [4]). The decay curves of this group in both sources were measured for about 15 years. (Fig. 7 (Fig 2. of Ref. [4])). In both cases the intensity first grew with time and then decayed with a half-life of 2.85 y. This group was identified as the decay of $^{236}$Pu from the energy and half-life of the α-particles. From the growth of the intensity at the beginning it was concluded that long-lived isomeric states were formed in the neutron deficient $^{236}$Am and $^{236}$Bk nuclei that decayed eventually to the g.s. of $^{236}$Pu.

d2) Isomeric states have been observed in chemically separated sources of Bk, Es and No-Lr, produced via secondary reactions where low energy α-particle groups (compared to g.s. to g.s. transitions) and five to seven orders of magnitude enhanced half-lives (compared to usual penetrability calculations) were seen [13]. These groups were consistently interpreted, both from the point of view of their low energies and their enhanced half-lives, as SD to SD and HD to HD α-transitions [13]. See for example Fig. 8 (Fig. 4 of Ref. [13]).

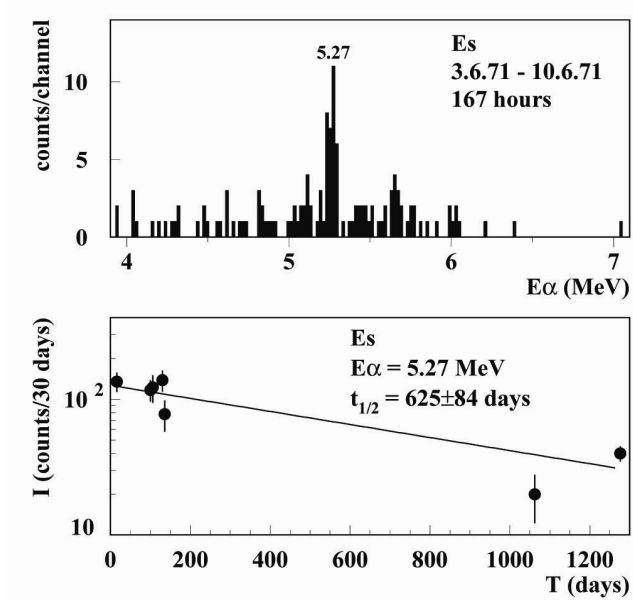

Fig. 8 (Fig. 4 of Ref. [13]). Top: α-particle spectrum obtained with the Es source produced by the secondary reaction experiments. Bottom: Decay curve of the relatively low energy 5.27 MeV α-particle group seen in the top figure.

d3) An isomeric state was seen in $^{210}$Fr produced via an ordinary $^{16}$O + $^{197}$Au heavy ion reaction [10]. A low energy and five orders of magnitude enhanced α-particle group of 5.2 MeV was found in coincidence with γ-ray transitions of a SD band. (See Table 2 (Table 2 of Ref. [10]) and Fig. 9.). They fit the rule $E_g = 4.40 \times J(J+1)$ keV and the parameter 4.40 keV is characteristic of SD band γ-rays. Based on coincidence measurements proving that the α-particles decayed to a SD band state, and penetrability calculations, it was deduced that the α-transition is from a long-lived high spin SD isomeric state in the parent nucleus to a high spin SD state in the daughter. In addition, long-lived proton radioactivities were found in this reaction [11]. Since the positions of the SD minima in $^{210}$Fr and nearby nuclei are predicted to be above the proton separation energies, it was hypothesized that the proton decays are due to transitions from long-lived SD isomeric state(s) in the parent(s) nuclei to normal states in the daughter(s).

Table 2 (Table 2 of Ref. [10]).
Energies of the γ-rays seen in coincidence with the 5.2 α-particles compared to transitions assuming $E_\gamma = 4.40 \times J(J+1)$.

| Transition | $E_\gamma$ (expt.)[a] (keV) | $E_\gamma$ (theor.) (keV) | $\Delta E$ (keV) |
|---|---|---|---|
| 3 ⇒ 2 | 26.8 | 26.4 | +0.4 |
| (2 ⇒ 0)[b] | | | |
| 4 ⇒ 3 | 35.1 | 35.2 | −0.1 |
| 5 ⇒ 4 | 43.6 | 44.0 | −0.4 |
| (3 ⇒ 1)[b] | | | |
| 8 ⇒ 7 | 70.8[c] | 70.4 | +0.4 |
| 12 ⇒ 11 | 105.3 | 105.6 | −0.3 |
| 13 ⇒ 12 | 114.4 | 114.4 | 0.0 |
| (7 ⇒ 5) | | | |
| 16 ⇒ 15 | 141.0 | 140.8 | +0.2 |
| 18 ⇒ 17 | 157.8 | 158.4 | −0.6 |
| 12 ⇒ 10 | 203.0 | 202.4 | +0.6 |
| 13 ⇒ 11 | 219.5 | 220.0 | −0.5 |

[a] The peak to total ratio was 100% up to about 120 keV and reduced gradually to 26% at 220 keV.
[b] Highly converted.
[c] Three events.

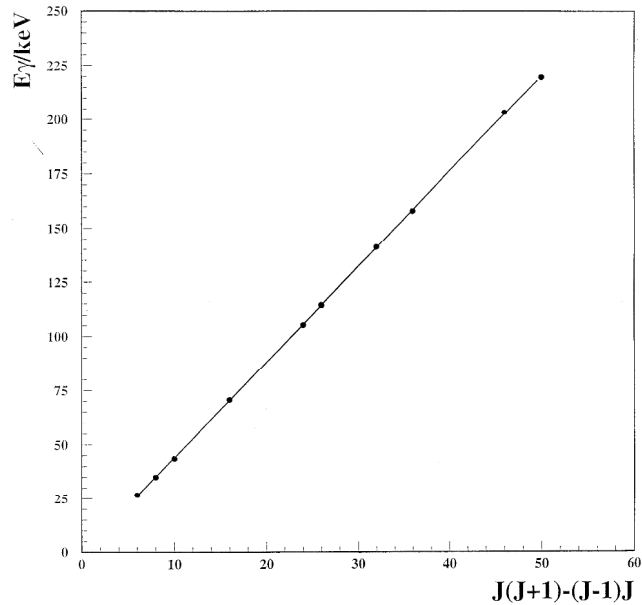

Fig. 9. Plot of $E_\gamma$ versus $J(J+1) - (J-1)J$ for the coincidence events given in Table 2.

d4) An isomeric state has been observed in $^{195}$Hg produced via an ordinary $^{28}$Si + $^{181}$Ta heavy-ion reaction [12]. A very high energy (8.6 MeV) and 13 orders of magnitude retarded α-particle group was seen in coincidence with SD band γ-ray transitions. (Fig. 1 (Fig. 5 of Ref. [12]) and Fig. 2 (Fig. 11 of Ref. [12])). (The $Q_\alpha$-value of the g.s. of this nucleus is 2.274 MeV so it does not decay by α-particles). The coincidence between the α-particles and the SD band γ-rays (Fig. 2 (Fig. 11 of Ref. [12])) proves that the α-particles decay to a SD state, and the high energy of the α-particles fits with predictions for α-decay of a HD state in the parent nucleus to a SD state in the daughter [12].

d5) Isomeric states have been observed in naturally-occurring Th substances by high resolution, low background Inductively Coupled Plasma-Sector Field Mass Spectrometer (ICP-SFMS) measurements, where the atomic masses were well separated from all molecules with the same mass number. Neutron-deficient $^{211,213,217,218}$Th isotopes, with half-lives $t_{1/2} \geq 10^8$ y, which is 16 to 22 orders of magnitude longer than their corresponding g.s., were seen [16]. See for instance Fig. 10 (Fig. 5 of Ref. [16]) for mass 218.

d6) In similar mass measurements to those described in d5), atoms with atomic mass numbers 261 and 265 with estimated half-lives of $t_{1/2} \geq 10^8$ y were found in natural Au substances. Because of the long lifetime it was deduced that they are isomeric states, most probably in $^{261}$Rg and $^{265}$Rg (eka-Au) [17]. (See for instance Fig. 11 (Fig. 2 of Ref. [17]) for mass 261.)

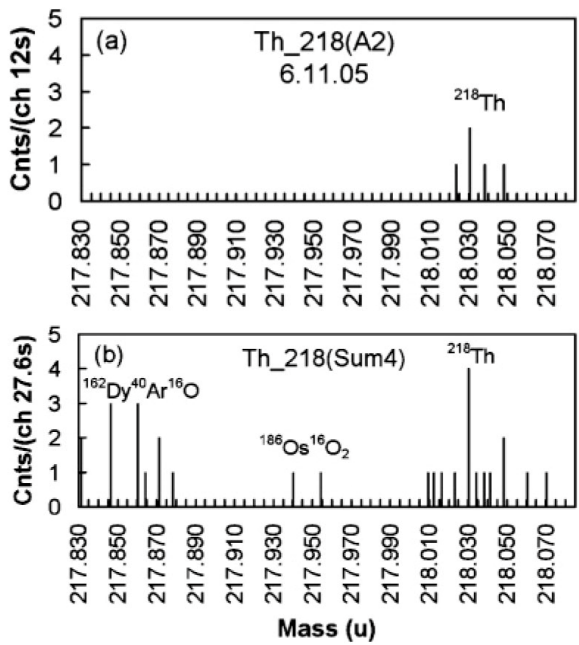

Fig. 10 (Fig. 5 of Ref.[16]. Copyright (2007) by the American Physical Society.) Results of measurements for mass region 218. (a) shows the spectrum obtained in run III, solution A, second measurement. The sum of four spectra is displayed in (b).

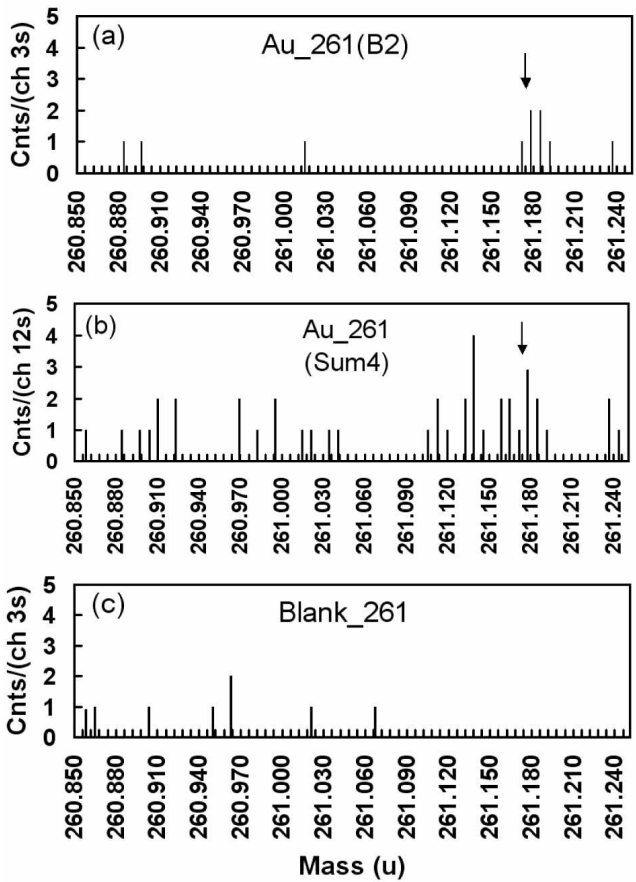

Fig. 11 (Fig. 2 of Ref [17]). Results of measurements for mass region 261. Figure 11(a) shows the results obtained in the second measurement with solution B. The sum of four spectra, two with solution A and two with solution B, is displayed in Figure 11(b). A spectrum of a blank solution is given in Figure 11(c). The arrows indicate the position of the predicted $^{261}$Rg mass.

d7) In high resolution mass measurements similar to those described in d5) and d6), a long-lived superactinide nucleus with A = 292, Z ≅ 122 (eka-Th) and $t_{1/2} \geq 10^8$ y, has been observed in natural Th. It should be in an isomeric state, and based on chemical arguments it is most probably an isomeric state in $^{292}$122 [18]. See Figs. 12 and 13 (Figs. 4 and 5 of Ref. [18]).

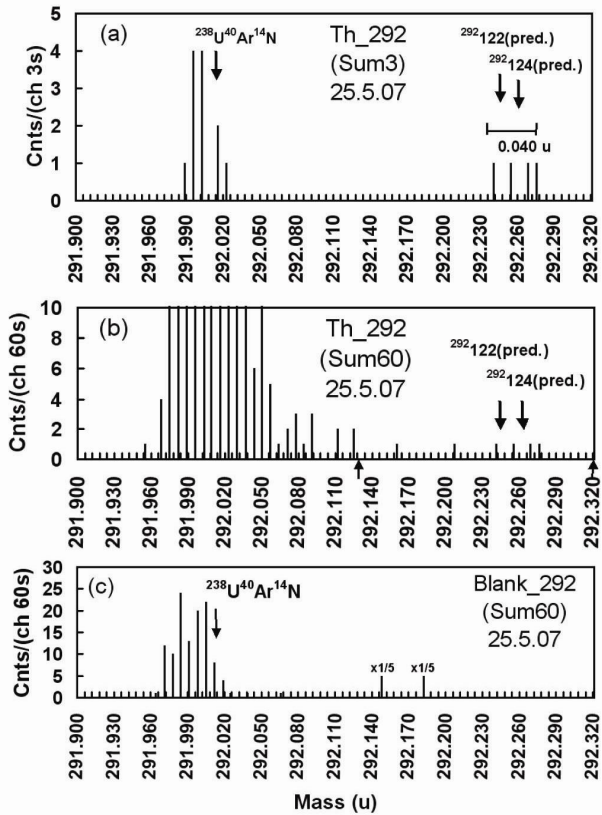

Fig. 12 (Fig. 4 of Ref. [18]). Results of measurements for mass region 292 obtained in the second run. The sum of three spectra out of the 60 measured is shown in (a), and the sum of the 60 spectra is shown in (b). The two rightmost arrows give the positions of the predicted masses of the $^{292}$122 [$^{292}$122(pred.)] and $^{292}$124 [$^{292}$124(pred.)] isotopes. The arrows in the left-hand parts of (a) and (c) show the known position of $^{238}$U$^{40}$Ar$^{14}$N. The c.m. of this peak is 292.003 u and its known mass is 292.016 u. The small upward-pointing arrows on the X-scale in (b) at masses 292.130 and 292.320 u indicate the limits chosen for the run of (c). In this region two background events are seen. If the four events of (a) were background counts, then about 80 counts should have been seen in this region in Fig. 12(b) instead of just two. The sum of 60 spectra of the blank solution is shown in (c).

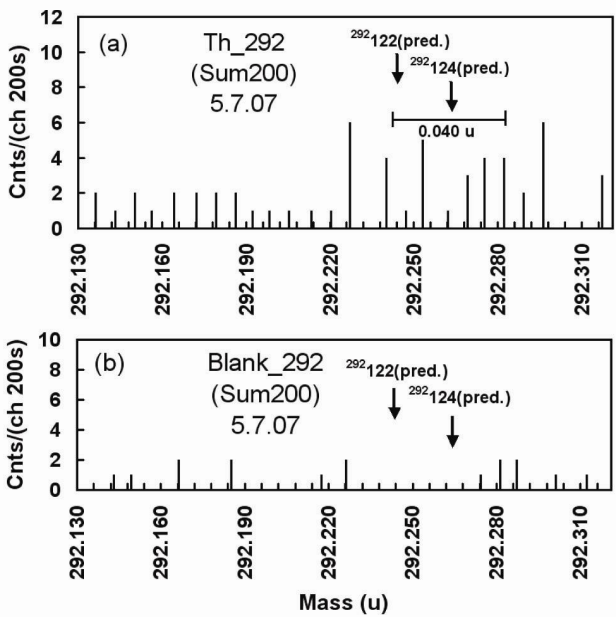

Fig. 13 (Fig. 5 of Ref. [18]). Results of measurements for mass region 292 obtained in the third run. The sum of 200 spectra in the limited mass region of 292.130 to 292.320 u (corresponding to the upward-pointing arrows on the X-scale in Fig. 12(b)) is shown in (a). Figure 13(b) shows the sum of 200 blank solution spectra. The arrows indicate the positions of the predicted masses of the $^{292}$122 [$^{292}$122(pred.)] and $^{292}$124 [$^{292}$124(pred.)] isotopes.

All the data presented in d1)-d6) were known to the JWP. As for d7) the preliminary results given in the conference contribution to TAN07 [18] were also known to the committee and a full preprint [18] was sent to them.

It is difficult to understand why such a redundancy of data that proves without a reasonable doubt the existence of long-lived isomeric states is ignored by the JWP without a single exception. It is well known that one factor that excites the nuclear community in the research for heavy elements is the theoretical predictions about the "island of stability". The possibility that nuclei find a way to stabilize themselves by up to 22 orders of magnitude compared to their normal g.s. decay, as reported in Ref. [16] (see d5)), and the prospects that this possibility opens for the future of heavy element research, where the results described in d5)-d7) are probably only the beginning, should, to our mind, please every nuclear scientist.

### e) COMPARISON WITH OTHER WORKS

The JWP Report states that "Results from other research groups that attempted obvious corroboration studies using multi-GeV protons incident on a U target clearly indicated the production path was irreproducible as previously stressed [1,3,4]. The latter refutations have been challenged by Brandt [18,25], a member of Marinov collaboration, in which he rejects the negative results because the same exact experiment as conducted by Marinov was not followed. However, *independent* evidence is what "Criteria" (*q.v.*) demand. Cloning of methodology is an approach that could easily camouflage systematic error."

The comparison of our experiments with other measurements was already dealt with back in 1984 [3] and it has recently been considered in some detail by Brandt [23], the scientist who at the time conducted most of those measurements. It was shown that:

e1) It is not correct that these experiments "clearly indicated the production path was irreproducible" since Ross *et al.* [24] "observed in the actinide sample definite evidence for unknown-activities in the energy range of (5.43-5.49) MeV just as the claims of the 'Marinov-Papers'[23]".

e2) The expected fission activity in their experiments [25,26] was $2 \times 10^{-3}$ fissions/day, 150 times lower than the background in their detector of 0.3 fissions/day. Thus, their negative results are not in contradiction with our positive results.

e3) The experiments with the U target [25,26] are irrelevant to the W target experiments as was shown in [3]. In order to get the isotope $^{272}$112 with a U target, the projectile should be $^{34}$Ca or a similar nucleus. The production cross sections of such neutron deficient nuclei in U are much lower compared to the production of for instance $^{88}$Sr in W. In addition, the $^{88}$Sr + $^{184}$W $\rightarrow$ $^{272}$112 reaction is quite cold and the $^{(34+x)}$Ca + $^{238}$U $\rightarrow$ $^{272}$112 + xn reactions are very hot with far more fission competition.

One finds nothing in these experiments that justify ignoring the positive evidence for the validity of the pathway producing heavy actinide nuclei like Am, Bk and Es as seen for example in Figs. 6-8. The irrelevance of the experiments mentioned in e2) and e3) is now recognized by Brandt [23] who headed the competing group that performed the experiments. It is not clear why the JWP sticks to this argument.

The JWP underestimates the importance of the rule that a repetition of an experiment should be a true repetition, by invoking the extraneous concept of "cloning", which is not in the "*Criteria*" [7]. If "cloning of methodology is an approach that could easily camouflage systematic error" then the two recognized events in α-chains are exactly cloning, as is the work of Morita.

### f) ACCURATE MASS MEASUREMENTS AND LONG–LIVED ISOMERIC STATES IN NATURALLY-OCCURING MATERIALS

The JWP ignores the results of accurate mass measurements that show the existence of long-lived isomeric states in the neutron-deficient $^{211,213,217,218}$Th nuclei found in natural Th [16], and in atomic masses 261 and 265 found in natural Au which are most probably $^{261}$Rg and $^{265}$Rg [17], as if they are due to unidentified species. "Such instruments are well known to generate ion species that are not readily identified."

The observed peaks at the known masses of the Th isotopes [16], and at the predicted masses of $^{261}$Rg and $^{265}$Rg or nearby superheavy elements [17], were found in very clean spectra. (See for example Figs. 10 and 11.) They are the result of repeated runs, with reproducible results, not a single scan. Furthermore, these peaks are well separated from all potential molecular-ion peaks of the same mass number. To disregard the results of 19 spectra measured in Ref. [16] and of eight spectra measured in Ref. [17] as "generated ion species" that all just happened to be at the known masses of $^{211,213,217,218}$Th and at the predicted masses of $^{261,265}$Rg is, to our mind, totally unjustified.

The JWP Report quotes the Element 2 manufacturer's brochure that "Even in a sample matrix as ultra pure water (UPW), interferences do exist, especially at low analyte concentration levels".

We did not claim in our papers that molecular interferences do not exist in our measurements, but rather that, because of the binding energy of the nuclei, their masses are different and well separated from the atomic masses found by us.

The JWP states that our "work does not convincingly demonstrate that the identification of the observed ions is unambiguous and correct" since we saw interferences like PbHe, AuO and DyArO.

There is nothing unusual in observing these molecules. Oxides are produced by interaction with the oxygen in the water solution, the ArO molecule is obtained from the oxygen in the solution and the Ar carrier gas, and the He molecule from impurity of He in the Ar carrier gas.

The JWP further claims that we "have relied on the manufacturer's specifications for (1) the relationship between sample concentration and ion current that should be observed as well as for

(2) the background trace peaks. In so doing, they calculate abundances of the trace peaks relative to the principal peak of $^{232}$Th of (1 to 10)x10$^{-11}$. Such an abundance ratio would require that the abundance sensitivity for this mass spectrometer exceeds that of any known mass spectrometer by at least two orders of magnitude and it is not accepted."

This argument has been raised by R. C. Barber and J. R. De Laeter [27] and has been refuted in [28]. The detection limit of an instrument depends on a) the basic performance of the instrument, the concentration of the solution and the duration of the measurement, and b) on the background in the measured region. The fact that the sensitivity in our experiment was suitable for measuring abundances in the region of $10^{-11}$–$10^{-12}$ can be seen from Fig. 1(a) of Ref. [16]. A peak due to 2 $\mu$g $l^{-1}$ of $^{209}$Bi with a total intensity of about 2.2x10$^7$ counts is seen in this figure. Normalization to 20 or 50 *mg l$^{-1}$* (concentrations that were used in the Au and Th isotope measurements) gives the abundance limit relative to Bi of 5x10$^{-12}$ and 2x10$^{-12}$, respectively. (The somewhat longer measuring periods for the $^{211,213,217,218}$Th isotopes of 24 to 30 s/ch compared to 18 s/ch for Bi reduces these values even further. In the case of Fig. 13 where the solution concentration was 80 *mg l$^{-1}$* and the measurement time was 200 s/ch, the limit of detection was 1x10$^{-13}$.) Such a procedure, where one estimates the intensity of the major peak at high concentration by normalization to data at low concentration, is standard in ICP-SFMS measurements. Whether or not the instrument is able to measure the current of the major peak is irrelevant, since the various concentrations are measured by weighing. In fact, the response of the ICP-SFMS is linear over 9-10 orders of magnitude. In the counting mode the maximum measurable counting rate is 5x10$^6$ counts/s. After that, the current is measured with the same detector up to an equivalent of 6x10$^9$ counts/s. The error in the cross-calibration between the counting and the analog mode is usually <2%. As regards the background in our measurements, it is not true that we relied on the manufacturer's specifications for the background trace peaks. Background spectra were measured by us as shown in Fig. 3(a) of Ref. [16] and in Figs. 11(c), 12(c) and 13(b). Since pure solutions were used in our measurements, the isotopes studied were far away from the relevant major peaks of $^{232}$Th or $^{197}$Au, and stable or long-lived radioactive isobars do not exist at the measured masses, it is seen that the background counts in the different measurements were very low. Hence, the newly observed peaks that fit the known masses of $^{211,213,217,218}$Th, the predicted masses of $^{261,265}$Rg and of A = 292, Z $\cong$ 122 nucleus, cannot be due to background.

Recently Lachner *et al.* [29] and Dellinger *et al.* [30] searched for the isomeric states in $^{211,213,217,218}$Th isotopes using AMS systems.[3] Dellinger *et al.* also searched for mass A = 292 in Th. Both groups report negative results. The upper limits of Ref. [29] for the ratios of $^{211,213,217,218}$Th/$^{232}$Th are in the region of 6.6x10$^{-13}$ - 2.4x10$^{-12}$, factors of 4.2 to 15.2 smaller than our smallest estimate of 1x10$^{-11}$. For the same limits Dellinger *et al.* [30] obtained values that are factors of 110 to 480 lower than the corresponding values of Lachner *et al.* [29], in the region of 5x10$^{-15}$. Regarding A = 292 the upper limit obtained by Dellinger *et al.* [30] is < 2x10$^{-15}$, which is 500 times smaller than our lower estimate of 1x10$^{-12}$.

A great advantage of the ICP-SFMS is its simplicity compared to the AMS system. The whole instrument is actually equivalent to the low energy part of the AMS. Even this part is much simpler in ICP-SFMS than in AMS, Positive ions are used in the ICP-SFMS. The selectivity between the different ions is not more than a factor of ten. On the other hand negative ion sputter source is used in AMS with large atomic and molecular selectivity. For instance, for Th solutions of 50 *mg l$^{-1}$* the typical ICP-SFMS input current is about 20 µA, or 400 times greater compared to 50 nA, which is the best $ThO_2^-$ current obtained by Middletown [31]. The transmission between the source and the detector in the ICP-SFMS is about $5\times10^{-5}$ for the medium resolution mode of *m/Dm* = 4000. The transmission between the low energy and the high energy regions in the AMS is about $1\times10^{-3}$ - $3\times10^{-3}$. Thus, the figure of merit for Th is about a factor of thirteen better for ICP-SFMS than for AMS. We do not think that discrepancies of factors of 4 to 15 between the upper limits of Lachner *et al.* [29] and our values [16], taking into account that different materials were used in the different experiments, justify rejection of our positive results, where some examples of them are shown in Fig. 10.

---

[3] These works were referred to by Barber and De Laeter in their *Note added in proof* [27].

Regarding the data of Dellinger *et al.* [30] where they report an upper limit 2000 times lower than our lower estimate, it is first not clear how they were able to obtain 100 to 500 times lower values compared to the values of Lachner *et al.* [29], both using AMS devices and basically the same starting material. The length of measurements was of comparable duration [29,32]. Let us add that Dellinger *et al.* [30] do see counts at the correct positions of $^{213,217}$Th, but they interpret them as pileup events. However, these counts did not appear where most of the pileup events should be, namely at the sum of two events that belong to the largest peak in the lower part of the energy spectrum [32].

Concerning the measurement of Dellinger *et al.* [30] on A = 292 where they got an upper limit 500 times lower than our lowest estimate, one can comment that the discrepancy mentioned above between the results of Dellinger *et al.* [30] and Lachner *et al.* [29] for the neutron deficient Th isotopes is also important here. In addition, it is not clear that setting the AMS for mass 292.0 u instead of 292.26 u [18] is correct. Thus, it is seen that there is no reason to ignore the positive results presented in Ref. [18] where examples of them are seen in Figs. 12 and 13.

g) FISSION, NUCLEOSYNTHESIS AND Po HALOS

The JWP states that "Species such as $^{210}$Th easily accessible through ordinary spallation, are expected to be very fissile."

The nuclei $^{210-218}$Th in their g.s. are known [33]. They decay by emitting α-particles and not by fission. Furthermore, they were produced by (HI,xn) reactions and not by spallation. The question of why the isomeric states seen by us do not decay by fission was dealt with in [14-16]. It is probably because of their high intrinsic spin. Such an effect has been predicted by Nilsson *et al.* [34], for instance, for an intrinsic $h_{11/2}$ state in N = 144−150 nuclei. The second barrier maximum increases by about 4 MeV, which corresponds to a 15 orders of magnitude increase in the fission half-life. Let us also mention that a high intrinsic spin state can be trapped between rotational band states, for example an intrinsic spin of $20\hbar$ between $18\hbar$ and $16\hbar$ rotational states, where E4 or M4 γ-rays are strongly inhibited.

The JWP further claims that "The natural nucleosynthetic path required for production of these nuclides is acknowledged by the Marinov collaboration to be inexplicable."

This statement is a misinterpretation of what we wrote. There is a difference between "inexplicable" and what is written in Ref. [16], "If the observed states turn out to be of the high spin SD and/or HD type, then heavy ion reactions could be involved in their nucleosynthesis." There is nothing unnatural in a process like heavy-ion reactions between low Z nuclei that form SD or HD states, which is followed by capture of slow or fast protons and neutrons. We might mention that the nucleosynthesis path for passing the gap from Pb to U nuclei is not entirely clear, yet this does not mean that anybody doubts the existence of U.

The JWP doubts the very existence of the Po halos and writes "Furthermore, advocacy by this collaboration of anomalous polonium halos (ostensibly from alpha-decay) in mica as additional evidence for long-lived superstates without referencing or acknowledging published, long-standing refutations, is misleading."

The question of the Po halos is a long-standing problem well known to the community. The work of Gentry (**Fossil alpha-recoil analysis of certain variant radioactive halos,** R. V. Gentry, Science **160**, 1228 (1968)) is quoted in our papers [14,15]. There can be no doubt about their existence (Fig. 14) which was dealt with by N. Feather [36]. From their measured radii it was deduced that they are due to α-decay of $^{210}$Po, $^{214}$Po and $^{218}$Po. These isotopes are short-lived and one does not see the rings from their predecessors when going up to long-lived $^{238}$U. Their existence was puzzling for many years. It is clear that long-lived isomeric states would furnish a natural explanation for them. This idea was suggested by Feather [36] "Obviously, if Po-replenishment of halo inclusions could be guaranteed once and for all by the initial incorporation of a long-lived parent (or parents), a model would be achieved which would avoid all the difficulties with which we have just been concerned."

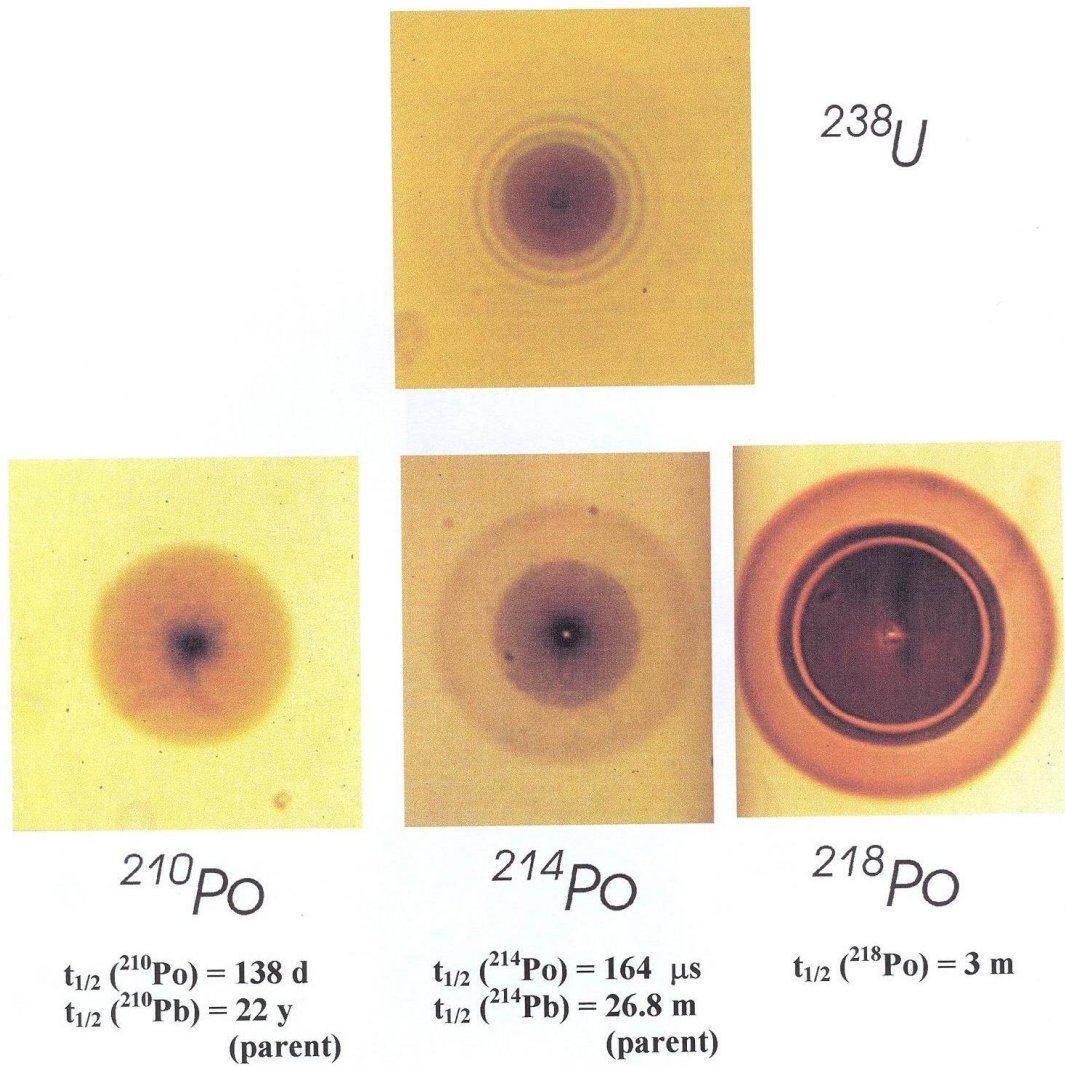

Fig. 14. Top: $^{238}$U halos. Rings are seen to have radii that fit the ranges of the α-particle groups of the $^{238}$U chain. They were formed as a result of the decay of a tiny U containing grain in the center of the halos over about $4.5 \times 10^9$ y.

Bottom, left: $^{210}$Po halo. One ring is seen with a radius that fits the range of the 5.3 MeV α-particle group from $^{210}$Po.

Bottom, center: $^{214}$Po halo. Two rings are observed with radii that fit the 7.6 MeV α-particle group from $^{214}$Po and the 5.3 MeV group from its daughter $^{210}$Po.

Bottom, right. $^{218}$Po halo. Three rings are seen with radii that fit the 6.0 MeV α-particle group from $^{218}$Po and the 7.6 and 5.3 MeV α-groups from its daughters $^{214}$Po and $^{210}$Po.

$^{210,214,218}$Po are short-lived and one does not see the rings due to any of their predecessors, going up to the long-lived parent $^{238}$U of the decay chain. (Pictures from Ref. [35]. Copyright (1986) by Robert V. Gentry.)

**g) Summary**

In summary it is seen that there is no reason for ignoring the discovery of element 112 back in 1971 and awarding the priority of its discovery to work done 25 years later in 1996..

**References**

[1] **Evidence for the Possible Existence of a Superheavy Element with Atomic Number 112,** A. Marinov, C. J. Batty, A. I. Kilvington, G. W. A. Newton, V. J. Robinson and J. H. Hemingway, Nature **229** (1971) 464-467.

[2] **Spontaneous fission previously observed in a mercury source,** A. Marinov, C. J. Batty, A. I. Kilvington, J. L. Weil, A. M. Friedman, G. W. A. Newton, V. J. Robinson, J. H. Hemingway and D. S. Mather, Nature **234** (1971) 212-215.

[3] **Consistent interpretation of the secondary-reaction experiments in W targets and prospects for production of superheavy elements in ordinary heavy-ion reactions,** A. Marinov, S. Eshhar, J. L. Weil and D. Kolb, Phys. Rev. Lett. **52** (1984) 2209-2212; **53** (1984) 1120 (E). http://prola.aps.org/abstract/PRL/v52/i25/p2209_1

[4] **Evidence for long-lived isomeric states in neutron-deficient $^{236}$Am and $^{236}$Bk nuclei,** A. Marinov, S. Eshhar and D. Kolb, Phys. Lett. B **191** (1987) 36-40.

[5] **Fusion of $^{16}$O + $^{148,150,152,154}$Sm at sub-barrier energies,** R. G. Stokstad, Y. Eisen, S. Kaplanis, D. Pelte, U. Smilansky and I. Tserruya, Phys. Rev. C **21** (1980) 2427-2435.

[6] **The evidence for production of the superheavy element with Z=112 via secondary and direct heavy ion reactions** A. Marinov, S. Gelberg and D. Kolb, Inst. Phys. Conf. Ser. No. 132, Sixth Int. Conf. *On Nuclei far from Stability* & Ninth Int. Conf. *On Atomic Masses and Fundamental Constants,* Bernkastel-Kues, Germany, Eds. R. Neugart and A. Wöhr (1992) 437-442.

[7] **Criteria that must be satisfied for the discovery of a new chemical element to be recognized. Part I,** A. H. Wapstra, Pure & Appl. Chem. **63** (1991) 879-886.

[8] **Discovery of the transfermium elements,** R. C. Barber, N. N. Greenwood, A. Z. Hrynkiewicz, Y. P. Jeannin, M. Lefort, M. Sakai, I. Ulehla, A. H. Wapstra and D. H. Wilkinson, Prog. Part. Nucl. Phys. **29,** 453-530 (1992).

[9] **Letter from Professor M. Lefort to A. Marinov of March 20, 1991**. Private communication.

[10] **Discovery of strongly enhanced low energy alpha decay of a long-lived isomeric state obtained in the $^{16}$O + $^{197}$Au reaction at 80 MeV, probably to superdeformed band,** A. Marinov, S. Gelberg and D. Kolb, Mod. Phys. Lett. A **11** (1996) 861-869. http://www.worldscinet.com/mpla

[11] **Evidence for long-lived proton decay not far from the b-stability valley produced by the $^{16}$O + $^{197}$Au reaction at 80 MeV,** A. Marinov, S. Gelberg and D. Kolb, Mod. Phys. Lett. A **11** (1996) 949-956.

[12] **Discovery of long-lived shape isomeric states which decay by strongly retarded high-energy particle radioactivity,** A. Marinov, S. Gelberg and D. Kolb, Int. J. Mod. Phys. E **10** (2001) 185-208. http://www.worldscinet.com/ijmpe

[13] **Strongly enhanced low energy alpha-particle decay in heavy actinide nuclei and long-lived superdeformed and hyperdeformed isomeric states,** A. Marinov, S. Gelberg and D. Kolb, Int. J. Mod. Phys. E **10** (2001) 209-236. http://www.worldscinet.com/ijmpe

[14] **New outlook on the possible existence of superheavy elements in nature,** A. Marinov, S. Gelberg, D. Kolb, R. Brandt and A. Pape, Phys. At. Nucl. **66** (2003) 1137-1145.

[15] **Coherent description for hitherto unexplained radioactivities by super- and hyperdeformed isomeric states,** A. Marinov, S. Gelberg, D. Kolb, R. Brandt and A. Pape, Int. J. Mod. Phys. E **12** (2003) 661-665.

[16] **Existence of long-lived isomeric states in naturally-occurring neutron-deficient Th isotopes,** A. Marinov, I. Rodushkin, Y. Kashiv, L. Halicz, , I. Segal, A. Pape, R. V. Gentry, H. W. Miller, D. Kolb and R. Brandt, Phys. Rev. C **76** (2007) 021303(R) 5 pages. http://scitation.aip.org/getabs/servlet/GetabsServlet?prog=normal&id=PRVCAN000076000002021303000001&idtype=cvips&gifs=yes

[17] **Existence of long-lived isotopes of a superheavy element in natural Au,** A. Marinov, I. Rodushkin, A. Pape, Y. Kashiv, D. Kolb, R. Brandt, R. V. Gentry, H. W. Miller, L. Halicz and I. Segal, Int. J. Mod. Phys. E **18** (2009) 621-629. http://www.worldscinet.com/ijmpe

[18] **Evidence for a long-lived superheavy nucleus with atomic mass number 292 and atomic number Z @ 122 in natural Th,** A. Marinov, I. Rodushkin, D. Kolb, A. Pape, Y. Kashiv, R. Brandt, R. V. Gentry and H. W. Miller, Third Int. Conf. on the Chemistry and Physics of the Transactinide Elements (TAN07), Davos, Switzerland, (2007) p. 65; arXiv:0804.3869; Accepted for publication in Int. J. Mod. Phys. E. http://www.worldscinet.com/ijmpe

[19] **On the discovery of the elements 110-112,** P. J. Karol, H. Nakahara, B. W. Petley and E. Vogt, Pure Appl. Chem. **73** (2001). 959-967.

[20] **On the claims for discovery of elements 110, 111, 112, 114, 116, and 118,** P. J. Karol, H. Nakahara, B. W. Petley and E. Vogt, Pure Appl. Chem. **75** (2003) 1601-1611.

[21] **Response to the IUPAC/IUPAP joint working party second report *'On the Discovery of Elements 110-118',*** A. Marinov, S. Gelberg, D. Kolb and G. W. A. Newton, nucl-ex/0411017. (Quoted in ENSDF (www.nndc.bnl.gov/superheavy.pdf).

[22] **Interaction of relativistic heavy ions in thick heavy element targets – and some unresolved problems,** R. Brandt *et al*. Phys. Part. Nucl. 39 (2008) 259-285.

[23] **Comments on the question of discovery of element 112 as early as 1971,** R. Brandt, Kerntechnik **70** (2005) 170-172.

[24] **Search for actinides in secondary reactions in tungsten and Au,** E. Ross, K. Bächmann, K.L Lieser, Z. Malek, D. Molzahn and J. Rudolf, J. Inorg. Nucl. Chem. **36** (1974) 251-257.

[25] **Search for superheavy elements in uranium irradiated with high-energy protons,** R. Brandt, D. Molzahn and P. Patzelt, Radiochimica Acta **18** (1972) 157-158.

[26] **Continued search for superheavy elements produced by secondary reactions at 24-GeV protons on uranium,** A. Boos**,** R. Brandt, D. Molzahn, P. Patzelt, P. Vater, K. Bächmann and E. Ross, J. Radiochemical and Radioanalytical Lett. **25** (1975) 357-364.

[27] **Comment on "Existence of long-lived isomeric states in naturally-occurring neutron-deficient Th isotopes",** R. C. Barber and J. R. De Laeter, Phys. Rev. C **79** (2009) 049801.

[28] **Reply to "Comment on 'Existence of long-lived isomeric states in naturally-occurring neutron-deficient Th isotopes' ",** A. Marinov, I. Rodushkin, Y. Kashiv, L. Halicz, I. Segal, A. Pape, R. V. Gentry, H. W. Miller, D. Kolb and R. Brandt, Phys. Rev. C **79** (2009) 049802.

[29] **Search for long-lived isomeric states in neutron-deficient thorium isotopes,** J. Lachner, I. I. Dillmann, T. Faestermann, G. Korschinek, M. Poutivtsev and G. Rugel, Phys. Rev. C **78**, 064313 (2008).

[30] **Search for a superheavy isotope with A=292 and neutron-deficient Th isotopes in natural thorianite,** F. Dellinger, O. Forstner, R. Golser, W. Kutschera, A. Priller, P. Steier, A. Wallner and G. Winkler, Poster presented at AMS 11, Rome, 2008.

[ 31] **A Cookbook of Negative Ions,** R. Middleton, Dept. of Phys., University of Pennsylvenia, 1990 (unpublished).

[32] **Searching for superheavy elements in thorianite (ThO$_2$),** W. Kutschera, F. Dellinger, O. Forstner, R. Golser, A. Priller, P. Steier, A. Wallner and G. Winkler, Talk given in Saraf Workshop, Maale Hachamisha, Israel, 2008.

[33] *Table of Isotopes*, edited by R. B. Firestone, V. S. Shirley, C. M. Baglin, S. Y. F. Chu and J. Ziplin (John Wiley and Sons, New York, 1996) and ENSDF.

[34] **On a new type of fission-isometric state,** S. G. Nilsson, G. Ohlén, C. Gustafsson and P. Möller, Phys. Lett. B **30**, 437 (1969).

[35] *Creation's Tiny Mystery,* R. V. Gentry, Earth Science Associates, Knoxville, TN (1986).

[36] **The unsolved problem of the Po-halos in Precambrian biotite and other old mineral,** N. Feather, Communications to the Royal Society of Edinburgh **11** (1978) 147-158.